\documentclass{cjpsuppl}% for LaTeX 2e
\usepackage{epsfig}
\begin{document}
\def\lapproxeq{\lower .7ex\hbox{$\;\stackrel{\textstyle <}{\sim}\;$}}
\def\gapproxeq{\lower .7ex\hbox{$\;\stackrel{\textstyle >}{\sim}\;$}}
\def\be{\begin{equation}}
\def\ee{\end{equation}}
\def\bea{\begin{eqnarray}}
\def\eea{\end{eqnarray}}
\def\ktbold{\mbox{\boldmath${k}$}_T}
\def\funp{{I\!\!P}}
\def\gtrsim{ \;\raisebox{-.7ex}{$\stackrel{\textstyle >}{\sim}$}\; }
\def\lesim{ \;\raisebox{-.7ex}{$\stackrel{\textstyle <}{\sim}$}\; }
\newcommand{\ksq}{k2} \newcommand{\qsq}{q2}
\newcommand{\epem}{e^+e^-}

\def\GeV{{\rm GeV}}
\def\MeV{{\rm MeV}}
\def\eV{{\rm eV}}
\def\ra{ \rightarrow }
\def\qq{{q\bar{q}}}
\def\bb{{b\bar{b}}}
\title{Diffractive Higgs production and related processes}
%\authori{A.B~Kaidalov}
%\addressi{Institute of Theoretical and Experimental Physics, Moscow, 117259, Russia}
\authori{A.D. Martin, A.B. Kaidalov, V.A.~Khoze, M.G. Ryskin and W.J. Stirling}    \addressi{Institute for Particle Physics Phenomenology,
Durham University, DH1 3LE, UK}
%\authorii{}    \addressii{Petersburg Nuclear Physics Institute, Gatchina,
%St.~Petersburg, 188300, Russia}
\authorii{}   \addressii{}
\authoriii{}     \addressiii{}
\authoriv{}   \addressiv{}
\authorv{}     \addressv{}
\authorvi{}    \addressvi{}
\headtitle{ {Diffractive Higgs production and related processes} }
\headauthor{A.D. Martin}
\lastevenhead{A.D.Martin \ldots Diffractive Higgs production \ldots}
%%%%%%%%%%%%%% Pro editory supplementu: %%%%%%%%%%%%%%%
\refnum{}%slouzi editorum pro evidenci; nakonec {}
\daterec{21 September 2004\\}
\suppl{A}  \year{2004} \setcounter{page}{1}
%\firstpage{1}
%\lastpage{000}
%\makefirsttitle
%%%%%%%%%%%%%%%%%%%%%%%%%%%%%%%%%%%%%%%%%%%%%%
\maketitle

%{\Large \bf Diffractive Higgs Production}
%
%\vspace*{1cm}
%\textsc{A.B~Kaidalov$^{a,b}$, V.A.~Khoze$^{a,c}$, A.D. Martin$^a$, and M.G. Ryskin$^{a,c}$}
%
%\vspace*{0.5cm} $^a$ Institute for
%Particle Physics Phenomenology,
%Durham University, DH1 3LE, UK \\[0.5ex]
%$^b$ Institute of Theoretical and Experimental Physics, Moscow, 117259, Russia\\[0.5ex]
%$^c$ Petersburg Nuclear Physics Institute, Gatchina,
%St.~Petersburg, 188300, Russia \\
%\end{center}
%
%\vspace*{1cm}
\begin{abstract}
We review the signal, and the $\bb$ background, for Higgs production by the
exclusive double-diffractive process, $pp \to p+H+p$, and its subsequent $H \to \bb$ decay, at the
LHC.  We discuss the production of Higgs bosons in both the SM and MSSM.  We show how the
predicted rates may be checked at the Tevatron by observing the exclusive double-diffractive
production of dijets, or $\chi_c$ or $\chi_b$ mesons, or $\gamma \gamma$ pairs.
\end{abstract}

\section{Introduction}

The identification of the Higgs boson(s) is one of the main goals
of the Large Hadron Collider (LHC) being built at CERN. There are
expectations that there exists a `light' Higgs boson with mass
$M_H\lapproxeq130$~GeV. In this mass range, its detection at the
LHC will be challenging. There is no obvious perfect detection
process, but rather a range of possibilities, none of which is
compelling on its own.  {\em Either}
large signals are accompanied by a huge background, {\em or} the
processes have comparable signal and background rates for which
the number of Higgs events is rather small.

Here we wish to draw attention to the exclusive signal $pp\ra p + H + p$,
where the +~sign indicates the presence of a rapidity gap. It may be
possible to install proton taggers so that the `missing mass' can
be measured very accurately. The experimental challenge is to provide
a set-up in which the bulk of the proton-tagged signal is deposited in a
small missing mass
window $\Delta M_{\rm missing}$
\cite{DKMOR}. The exclusive process allows the mass of
the Higgs to be measured in two independent ways. First the tagged
protons give $M_H = M_{\rm missing}$ and second, via the $H\ra\bb$
decay, we have $M_H = M_{\bb}$, although now the resolution is
much poorer with $\Delta M_{\bb}\simeq10$~GeV or more. The existence of
matching peaks, centered about $M_{\rm missing}=M_{\bb}$, is a
unique feature of the exclusive diffractive Higgs signal. Besides
its obvious value in identifying the Higgs, the mass equality also
plays a key role in reducing background contributions. Another crucial
advantage of the exclusive process $pp\ra p+H+p$, with $H\ra\bb$,
is that the leading order $gg\ra\bb$ background subprocess is
suppressed by a $J_z=0$, P-even selection rule \cite{KMRmm,DKMOR}.

\section{Calculation of the exclusive Higgs signal}

\begin{figure}
\begin{center}
\centerline{\epsfxsize=0.4\textwidth\epsfbox{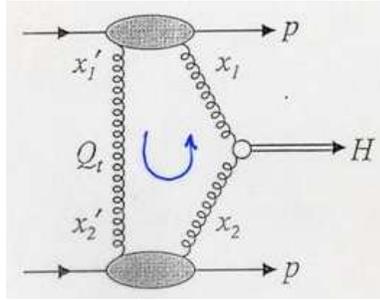}}
\caption{Schematic diagram for exclusive Higgs production at the LHC,
$pp \to p+H+p$. The presence of Sudakov form factors ensures the infrared
stability of the $Q_t$ integral over the gluon loop. It is also necessary
to compute the probability, $S^2$, that the rapidity gaps survive soft rescattering.}
\label{fig:H}
\end{center}
\end{figure}

The basic mechanism for the exclusive process, $pp\ra p+H+p$, is
shown in Fig.~$\ref{fig:H}$. Since the dominant contribution comes from the
region $\Lambda_{\rm QCD}^2\ll Q_t^2\ll M_H^2$ the amplitude may
be calculated using perturbative QCD techniques\cite{KMR,KMRmm}
\begin{equation}
{\cal M}_H \simeq N\int\frac{dQ^2_t\ V_H}{Q^6_t}\: f_g(x_1, x_1', Q_t^2, \mu^2)f_g(x_2,x_2',Q_t^2,\mu^2), \label{eq:M}
\end{equation}
where the overall normalization constant $N$ can be written in terms of the $H\to gg$ decay width\cite{INC}, and where the
$gg\to H$ vertex factors for CP $=\pm 1$ Higgs production are, after azimuthal-averaging,
\be
V_{H(0^+)} \simeq  Q_t^2, ~~{\rm and}~~
V_{A(0^-)}  \simeq  (\vec p_{1t} \times \vec p_{2t})\cdot \vec n_0,
\label{eq:rat4}
\ee
Expressions (\ref{eq:M},\ref{eq:rat4}) hold for small $p_{it}$, where the $\vec p_{it}$
are the transverse momenta of the outgoing protons,
and $\vec n_0$ is a unit vector in the beam direction.
The $f_g$'s are the skewed unintegrated gluon densities at the hard scale $\mu$,
taken to be $M_H/2$. Since $(x'\sim Q_t/\sqrt s)\ll (x\sim M_H/\sqrt s)\ll 1$, it is possible
to express $f_g(x,x',Q_t^2,\mu^2)$, to single log accuracy, in
terms of the conventional integrated density $g(x)$. The $f_g$'s
embody a Sudakov suppression factor $T$, which ensures that the gluon does not radiate in the
evolution from $Q_t$ up to the hard scale $M_H/2$, and
so preserves the rapidity gaps.  The apparent infrared divergence of~(\ref{eq:M}) is nullified
 for $H(0^+)$ production by these Sudakov factors.
 However the amplitude for $A(0^-)$ production is much more sensitive to the infrared contribution. Indeed
let us consider the case of small $p_{it}$ of the outgoing protons. Then we see, from~(\ref{eq:rat4}), that
 the $dQ^2_t/Q_t^4$ integration for $H(0^+)$ is
replaced by $p_{1t}p_{2t} dQ^2_t/Q_t^6$ for $A(0^-)$, and now the Sudakov suppression is not
enough to prevent a significant contribution from the $Q_t^2\lesim1~\GeV^2$ domain.

The radiation associated with the $gg\ra H$ hard
subprocess is not the only way to populate and to destroy the
rapidity gaps. There is also the possibility of soft rescattering
in which particles from the underlying event populate the gaps.
The probability, $S^2=0.026$ at the LHC, that the gaps survive the soft
rescattering was calculated using a two-channel eikonal model,
which incorporates high mass diffraction\cite{KMRsoft}. Including
this factor, and the NLO $K$ factor, the cross section is
predicted to be \cite{INC}
\begin{equation}
\sigma(pp\ra p+H+p)\simeq 3\:{\rm fb} \label{eq:sigma}
\end{equation}
for the production of a Standard Model Higgs boson of mass 120~GeV
at the LHC. It is
estimated that there may be a factor of 2.5 uncertainty (up or down) in this
prediction\cite{KKMR4}.

If we include a factor 0.6 for
the efficiency associated with proton tagging, 0.67 for the $H\ra\bb$ branching fraction,
0.6 for $b$ and
$\bar{b}$ tagging, 0.5 for the $b,\bar{b}$ jet polar angle cut,
$60^\circ<\theta<120^\circ$, (necessary to reduce the $\bb$ QCD
background)\cite{DKMOR}, then,
for a luminosity of ${\cal L}=30\:{\rm fb}^{-1}$, the original $3 \times 30=90$
 events are reduced to an observable
signal of 11 events.

\section{Background to the exclusive Higgs signal}

The advantage of the $p+(H\ra\bb)+p\,$ signal is that there exists
a $J_z=0$ selection rule, which requires the leading order
$gg^{PP}\ra\bb$ background subprocess to vanish in the limit of
massless quarks and forward outgoing protons. (The $PP$ superscript is
to note that each gluon comes from colour-singlet $gg~t$-channel exchange.) However,
in practice, LO background contributions remain. The prolific
$gg^{PP}\ra gg$ subprocess may mimic $\bb$ production since we may
misidentify the outgoing gluons as $b$ and $\bar{b}$ jets.
Assuming the expected 1\% probability of misidentification, and
applying $60^\circ<\theta<120^\circ$ jet cut, gives a
background-to-signal ratio $B/S \sim 0.06$. (Here, for reference, we
assume that the bulk of the Higgs signal can be collected within
an interval $\Delta M_{\rm missing}=1$ GeV.) Secondly, there is an
admixture of $|J_z|=2$ production, arising from non-forward going
protons which gives $B/S \sim 0.08$. Thirdly, for a massive quark
there is a contribution to the $J_z=0$ cross section of order
$m_b^2/E_T^2$, leading to $B/S \sim 0.06$, where $E_T$ is the
transverse energy of the $b$ and $\bar{b}$ jets.

Next, we have the possibility of NLO $gg^{PP}\ra\bb g$ background
contributions. Of course, the extra gluon may be observed
experimentally and these background events eliminated. However,
there are exceptions. The extra gluon may go unobserved in the
direction of a forward proton. This background may be effectively
eliminated by requiring the equality $M_{\rm missing} = M_{\bb}$.
Moreover, soft gluon emissions from the initial $gg^{PP}$ state
factorize and, due to the overriding $J_z=0$ selection rule, these
contributions to the QCD
$\bb$ production are also suppressed. The remaining danger is
large angle hard gluon emission which is collinear with either the
$b$ or $\bar{b}$ jet, and therefore unobservable. If the cone
angle needed to separate the $g$ jet from the $b$ (or $\bar{b}$)
jet is $\Delta R \sim 0.5$ then the expected background from
unresolved three jet events leads to $B/S \simeq 0.06$.
The NNLO $\bb gg$ background contributions are found to be
negligible (after requiring $M_{\rm missing}\simeq M_{\bb}$), as
are soft Pomeron-Pomeron fusion contributions to the background
(and to the signal)~\cite{DKMOR}.  Also note that radiation off the
screening gluon, in Fig.~$\ref{fig:H}$, is numerically small\cite{myths}.

\section{The signal-to-background ratio}

So, in total, for the exclusive production of a 120 GeV (SM) Higgs boson at the LHC with
the integrated luminosity ${\cal L}=30$ fb$^{-1}$,
the signal-to-background ratio is
\be
S/B~\simeq~(1 \GeV/\Delta M_{\rm missing}) ~11/4 ~~{\rm events},
\ee
after cuts and acceptance. This corresponds to a statistical significance of roughly
$3.7\sigma~\sqrt(1 \GeV/\Delta M_{\rm missing})$.
That is, if almost the whole Higgs signal can be collected within the
interval $\Delta M_{\rm missing}=1$ GeV, then $S/B \simeq 3$, corresponding to a $3.7\sigma$ signal.
In the case of a Gaussian missing mass distribution of
width $\sigma$, about 87\% of the signal is contained in a bin
$\Delta M_{\rm missing}=3\sigma$, that is $M_{\rm missing}=M_H \pm 1.5\sigma$.

We could consider Higgs production in other diffractive channels, such as diffractive
production accompanied by
proton dissociation ($pp \to M_1+H+M_2$), or central inelastic production
($pp \to p+(M \to HX)+p$).  However they are worse than the usual totally
inclusive production --
there is no precise missing mass measurement, no selection rule to suppress the
background and more serious pile-up problems.  The somewhat smaller density of soft secondary hadrons in the
Higgs rapidity region does not compensate for the much smaller statistics
(cross sections) in diffractive processes.

\section{Exclusive SUSY Higgs signals}

\begin{figure}
\begin{center}
\centerline{\epsfxsize=0.8\textwidth\epsfbox{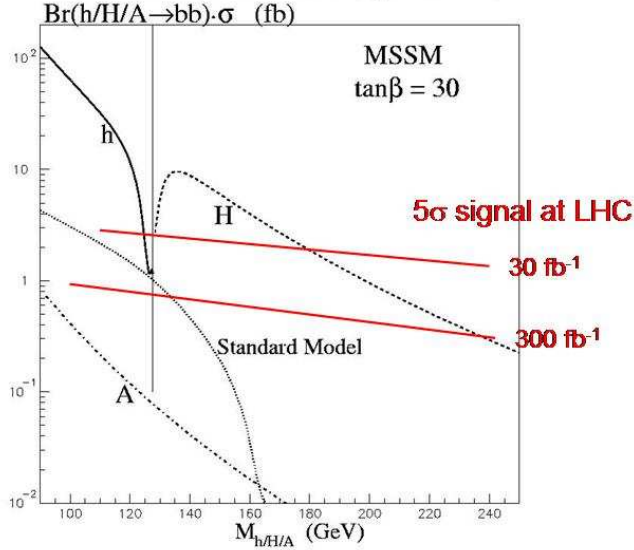}}
\caption{The cross sections predicted for the exclusive diffactive
production of $h(0^+), H(0^+)$ and $A(0^-)$ MSSM Higgs bosons at the LHC, for
tan$\beta =30$.   The superimposed lines show the cross section
required for a $5\sigma$ signal for integrated LHC luminosities of
30 and 300 ${\rm fb}^{-1}$. The figure is taken from Ref.~\cite{KKMR4}.}
\label{fig:SUSY}
\end{center}
\end{figure}

To be specific, we discuss the three neutral Higgs bosons of the MSSM
model: $h,H$ with CP=1 and $A$ with CP=--1.  There are regions of MSSM
parameter space where the conventional signals ($\gamma\gamma, WW, ZZ$
decays) are suppressed, but where the exclusive subprocess $gg\ra H \ra \bb$ is strongly
enhanced\cite{KKMR4}.  For example, for $M_A$ = 130 GeV and tan$\beta$ = 50, we have
$M_h$ = 124.4 GeV with $S/B=71/3$ events, $M_H$ = 135.5 GeV with $S/B=124/2$
events and $M_A$ = 130 GeV with $S/B=1/2$ events, so both
$h$ and $H$ should be clearly visible. (Again, for reference,
we assume that $\Delta M_{\rm missing}=1$ GeV can be achieved.)  The decoupling regime
($M_A \gtrsim 2M_Z$ and tan$\beta \gtrsim 5$) is another example where the exclusive
signal is of great value.   In this case $h$ is indistinguishable from a SM Higgs, and so the discovery
of $H$ is crucial to establish the underlying dynamics.  The plot of Fig.~$\ref{fig:SUSY}$,
with tan$\beta =30$, shows that a $5\sigma$ signal is possible up to quite large values of $M_H$.

If the exclusive cross sections for scalar and pseudoscalar Higgs production were comparable,
it would be possible to separate them readily by a missing mass scan, and by the study of
azimuthal correlations between the outgoing protons.  Unfortunately
pseudoscalar exclusive production is strongly suppressed by the P-even selection.
Maybe the best chance to identify the $A(0^-)$ boson is through the double-diffractive
process, $pp\ra X+A+Y$, where both
protons dissociate\cite{KKMR4}.

\section{Related processes: checks of the predicted exclusive Higgs yield}

The exclusive Higgs signal is particularly clean, and the signal-to-background
ratio is especially favourable, in comparison with the other proposed
detection modes.  However the expected number of events is low.
Therefore it is important to check the predictions for exclusive Higgs production
by studying processes mediated by the same mechanism, but
with rates which are sufficiently high that they may be observed at the Tevatron
(as well as at the LHC).  The most obvious examples are those in which the Higgs of
Fig.~$\ref{fig:H}$ is replaced by either a dijet system, a $\chi_c$ or $\chi_b$ meson, or
by a $\gamma \gamma$ pair.

First, we discuss the exclusive production
of a pair of high $E_T$ jets, $p\bar {p} \to p+jj+\bar {p}$ \cite{KMR,INC}.
This would provide an effective $gg^{PP}$ `luminosity
monitor' just in the kinematical region of the Higgs
production. The corresponding cross section was evaluated to
be about 10$^4$ times larger than that for the SM Higgs boson.
Thus, in principle,
this  process appears to be an ideal `standard candle'.  The expected cross section is rather large,
and we can study its behaviour as a function of the mass of the dijet
system.  Unfortunately, in the present CDF environment, the
background from `inelastic Pomeron-Pomeron collisions' is large as well.
Theoretically the exclusive dijets should be observed as a narrow peak,
sitting well above the background, in the
distribution of the ratio
\be
R_{jj}=E_{{\rm dijet}}/E_{\rm {PP}}
\ee
 at $R_{jj}=1$, where $E_{\rm {PP}}$ is the energy of the incoming
 Pomeron-Pomeron system.  In practice
the peak is smeared out due to hadronization and the jet-searching algorithm.
For jets with $E_T=10$ GeV and a jet cone $R<0.7$, more than 1 GeV will be lost
outside the cone, leading to (i) a decrease of the measured jet energy of about 1-2 GeV,
and, (ii) a rather wide peak ($\Delta R_{jj}\sim \pm 0.1$) in the $R_{jj}$ distribution.
The estimates based on Ref.~\cite{INC} give an exclusive cross section for dijet
production with $E_T>25$ GeV (and CDF cuts) of about 40 pb, which is very close
to the recent CDF measurement\cite{CDFchi},
\be
\sigma(R_{jj}>0.8,~E_T>25 ~{\rm GeV})~~=~~34~\pm 5({\rm stat}) ~\pm 10({\rm syst}) ~{\rm pb}.
\ee
 However there is no `visible' peak in the CDF data for $R_{jj}$ close to 1. The contribution
 from other channels (called Central Inelastic in Ref.~\cite{INC}) is too large, and
  matches with the expected peak smoothly\footnote{We hope that applying the $k_t$
  jet searching algorithm, rather than the jet cone algorithm, would improve the
 selection of the exclusive events.   This is in accord with the studies in Ref.~\cite{CFP}.}.

\begin{figure}
\begin{center}
\centerline{\epsfxsize=0.8\textwidth\epsfbox{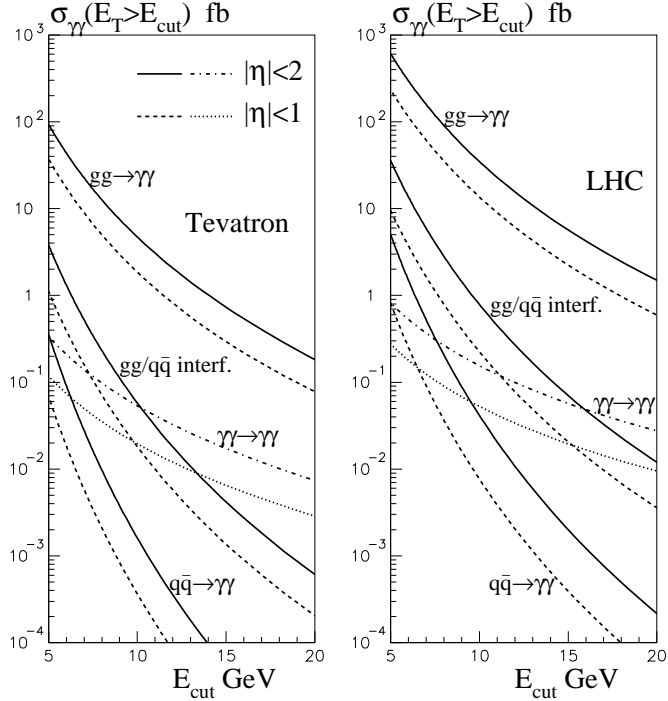}}
\caption{The contributions to the cross section for exclusive $\gamma \gamma$ production
from $gg$ and $\qq$ exchange at the Tevatron and the LHC.   Also shown is the
contribution from the QED subprocess $\gamma \gamma \to \gamma \gamma$.  For each component we
show the cross section restricting the emitted photons to have $E_T>E_{\rm cut}$
and to lie in the centre-of-mass
rapidity interval $|\eta_{\gamma}|<1$ (or $|\eta_{\gamma}|<2$).
The figure is taken from Ref.~\cite{KMRSgam}.}
\label{fig:gamma}
\end{center}
\end{figure}

An alternative possibility is to measure exclusive double-diffractive $\gamma\gamma$ production
with high $E_T$ photons, that is $p\bar {p} \to p+\gamma \gamma +\bar {p}$ \cite{INC,KMRSgam}.
Here there are no problems with hadronization or with the identification of the jets.
On the other hand the exclusive cross section is rather small. As usual, the
perturbative QCD Pomeron is described by two (Reggeized) gluon exchange.
 However the photons cannot be emitted from the gluon lines
directly. We need first to create quarks. Thus a quark loop is required, which causes an extra coupling
 $\alpha_s(E_T)$ in the amplitude.
The predictions of the cross section for
 exclusive $\gamma \gamma$ production are shown in Fig.~$\ref{fig:gamma}$.

Recently the first `preliminary' result on exclusive $\chi_c$ production has been
reported\cite{CDFchi}. Although it is consistent with perturbative QCD expectations\cite{KMRSchi},
the mass of the $\chi_c$-boson, which drives the scale of the process, is too
low to justify just the use of perturbative QCD\footnote{Even lower scales
correspond to the fixed target central double diffractive meson resonance production
observed by the WA102 collaboration at CERN\cite{WA102}.  Therefore,
it is intriguing that the qualitative features of the observed $p_t$ and
azimuthal angular distributions appear to be in good agreement with the
perturbatively based expectations\cite{KMRtag}.}. However,
in Ref.~\cite{KMRSchi}, it was found that both a Regge formalism and perturbative QCD predict essentially the
same qualitative behaviour for the central double-diffractive production of
`heavy' $\chi_c(0^{++})$ and $\chi_b(0^{++})$ mesons. Due to the
low scale, $M_\chi /2$, there is a relatively small contribution coming from the process
in which the incoming protons dissociate. Therefore simply
selecting events with a rapidity gap on either side of the $\chi$, almost
ensures that they will come from the exclusive reaction, $p\bar {p} \to p\ +\ \chi\ +\ \bar {p}$.
Although exclusive $\chi$ production is expected to dominate, the  predicted\cite{KMRSchi} event rates are large
enough to select double-diffractive dissociative events with large transverse
energy flows in the proton fragmentation regions. Such events are particularly
interesting.  First, in this case, the large value of
$E_T$ provides the scale to justify the validity, and the reasonable
accuracy, of the perturbative QCD calculation of the cross section. Next, by measuring the
azimuthal distribution between the two $E_T$ flows, the
parity of the centrally produced system can be determined.

Another possible probe of the exclusive double-diffractive
formalism would be to observe central open $\bb$ production;
namely $b,{\bar b}$ jets with $p_t \gapproxeq m_b$.
Again, this would put the application of perturbative QCD on a sounder footing.
It would allow a check of the perturbative formalism, as well as
a study of the dynamics of $\bb$ production.

\section{Conclusion}

If the Higgs is light, $M_H \lapproxeq 135$ GeV, it will be experimentally
challenging to study it in detail at the LHC.   All possible processes should
be considered.   Here we have emphasised the unique advantages of exclusive
double-diffractive Higgs production, {\it provided} the forward
outgoing protons can be precisely tagged.   The missing mass, $M_{\rm missing}$, measured by
the forward proton detectors can then be matched with the mass $M_{\bb}$ from the main decay
mode, $H \to \bb$.  Moreover the QCD $\bb$ background is suppressed by a $J_z=0$ selection rule.
The events are clean, but the predicted yield is low: about 10 events,
after cuts and acceptance, for an integrated luminosity of
${\cal L}=30~{\rm fb}^{-1}$.  The signal-to-background ratio is about 1 or better,
depending crucially on the accuracy with which $M_{\rm missing}$ can be measured.
We have emphasized the importance of checking these perturbative QCD predictions by
observing analogous double-diffractive processes, with larger cross sections, at the
Tevatron.

\section*{Acknowledgements}

ADM thanks the Leverhulme Trust for an Emeritus Fellowship.  This work was supported by
the UK PPARC, by a Royal Society FSU grant, by grant RFBR 04-02-16073.

\end{document}